\documentclass[twocolumn,nofootinbib,superscriptaddress]{revtex4-1}

\pdfoutput=1

\usepackage{slashed}

\usepackage[normalem]{ulem}

\usepackage{color}

\usepackage{epsfig}
\usepackage{epstopdf}
\usepackage{epsf}
\input{epsf.sty}
\usepackage{graphicx,amsmath,amssymb}
\usepackage{epsfig}
\allowdisplaybreaks

\def\ei{\end{itemize}}
\def\be{\begin{equation}}
\def\ee{\end{equation}}
\newcommand{\bea}{\begin{eqnarray}}
\newcommand{\eea}{\end{eqnarray}}

\usepackage{bbm,bm,amsmath,amssymb}

\usepackage{hyperref}

\def\K{{K\"{a}hler}}
\def\Kahler{K\"{a}hler~}

\newcommand{\rf}[1]{(\ref{#1})}
\usepackage{float}

\begin{document}

\title{de Sitter Vacua with a Nilpotent Superfield}
\author{Renata Kallosh}
\email{kallosh@stanford.edu}
\affiliation{Stanford Institute for Theoretical Physics and Department of Physics, Stanford University, Stanford,
CA 94305, USA}
\author{Andrei Linde}
\email{alinde@stanford.edu}
\affiliation{Stanford Institute for Theoretical Physics and Department of Physics, Stanford University, Stanford,
CA 94305, USA}
\author{Evan McDonough}
\email{evan\_mcdonough@brown.edu}
\affiliation{Department of Physics, Brown University, Providence, RI, USA. 02903}
\author{Marco Scalisi}
\email{marco.scalisi@kuleuven.be}
\affiliation{Institute for Theoretical Physics, KU Leuven, Celestijnenlaan 200D, B-3001 Leuven, Belgium}

 \begin{abstract}

We study the arguments given in \cite{Moritz:2017xto} which suggest that the uplifting procedure in the KKLT construction is not valid. First we show that  the modification of the SUSY breaking sector of the nilpotent superfield, as proposed in \cite{Moritz:2017xto}, is not consistent with non-linearly realized local supersymmetry of de Sitter supergravity. Keeping this issue aside, we also show that the corresponding bosonic potential does actually describe de Sitter uplifting.

 \end{abstract}

\maketitle

%\vspace{10mm}

\section{Introduction}

The quest for four-dimensional de Sitter solutions to ten-dimensional string theory is a long-standing problem in high-energy theoretical physics. While many proposals exist \cite{Kachru:2003aw,Balasubramanian:2005zx},  this continues to be a subject of  some  debate.  In \cite{Polchinski:2015bea} it was argued that this task can be reduced to the identification of the correct four-dimensional effective field theory, while compatibility with known approaches to the stabilization of string theory moduli \cite{Kachru:2003aw,Balasubramanian:2005zx,Giddings:2001yu} demands the use of supergravity. One is then naturally led to the study of de Sitter solutions to $\mathcal{N}=1$ $d=4$ supergravity as a proxy for de Sitter solutions in full string theory.

One of the known examples of de Sitter (dS) in string theory is the KKLT construction \cite{Kachru:2003aw} where the `uplift' from AdS to dS  is provided by an anti-D3 brane. In the language of a 4d effective action, it has been argued \cite{Kallosh:2014wsa,Bergshoeff:2015jxa} that this scenario  can be  described by\footnote{The ``warped'' case $K= -3\log\left(T+\bar T -S \bar S\right)$ gives a different \mbox{$T$-dependence} in the uplifting term and very little changes in terms of the conclusions of the present paper.}
\be
\begin{aligned}\label{KKLT}
&K= -3\log\left(T+\bar T\right)+S \bar S \, , \\
 &W= W_0 + A \exp(-a T) + b\ S\,,
\end{aligned}
\ee
where $T$ is the volume \Kahler modulus and $S$ is a nilpotent chiral superfield (i.e. $S^2=0$). The $S$ multiplet contains no fundamental scalar and its single degree of freedom may be identified with an $\overline{\rm D3}$ worldvolume fermion \cite{Kallosh:2014wsa,Bergshoeff:2015jxa}. Its fundamental role in the construction of cosmological models with positive vacuum energies has been repeatedly pointed out and extensively investigated (see e.g. \cite{Ferrara:2014kva,Kallosh:2014via,DallAgata:2014qsj,Kallosh:2014hxa,Scalisi:2015qga,McDonough:2016der,Kallosh:2017wnt,Dalianis:2017okk}). Furthermore, the coefficient $b$ encodes important information of the $\overline{\rm D3}$ brane and it is often represented as $b\equiv e^{2 {\cal {A}}_{0} } \mu^{2}$, where $e^{{\cal {A}}_{0}} \ll 1$ is the warp factor at the tip of the throat and $\mu$ is related to the unwarped tension of the  anti-brane as $|\mu^{4}| \sim T_{3}$.

It has been recently suggested in \cite{Moritz:2017xto} that a 10d analysis might reveal backreaction issues, which would ultimately spoil the possibility of realizing a positive cosmological constant in the KKLT construction. In the same article, the authors speculate that a 4d signal of this situation might be found in a coupling between the fields $T$ and $S$. 

While this coupling is absent in the original KKLT proposal, the  proposed reason to consider it was a  similarity to the case of D3 branes. It was suggested in  \cite{Baumann:2006th} that the coefficient of the non-perturbative superpotential $A$ might have a dependence on the modulus $\Phi$ describing a position of the ${\rm D3}$ brane, so that $A= A(\Phi)$. 
Based on this analogy, the authors of \cite{Moritz:2017xto} suggested that one can expect a dependence on the $\overline{\rm D3}$ brane so that $A$ becomes replaced by $ A+cS$.    
The resulting effective d=4 supergravity theory is of the following form   
\be
\begin{aligned}
\label{KWmix}
& K= -3\log\left(T+\bar T -S \bar S\right) \, , \\
& W= W_0 + A(1 + c\ S) \exp(-a T) + b\ S\,,
\end{aligned}
\ee
where higher order terms in $S$ vanish identically because of the assumed nilpotency condition.  The coefficient $c$ was suggested to be viewed as some  function of the anti-brane tension. 

It was argued in \cite{Moritz:2017xto} that  we do not know much about the value of the parameter $c$ and  one may expect it to be O(1), whereas $b$ is suppressed by a warping factor. Given this hierarchy $b\ll A\, c$, it was assumed that one can effectively ignore $b$ in investigations of the KKLT uplifting.  Then, one can show that taking $b=0$ does not lead to a dS vacuum state, for any value of $c$. 

However, before studying the consequences of this assumption, we would like to notice immediately that the argument that $b\ll A\, c$  because $c$ is not suppressed by the warping factor $e^{2 {\cal {A}}_{0} } \ll 1$ seems not well motivated. Indeed, consider the limiting case of super-strong warping, which describes the limit $b \to 0$, for fixed value of $T_{3}$. In this limit, the effective $\overline{\rm D3}$ brane stress tensor, as seen from the bulk, vanishes and, therefore, cannot cause any backreaction. In other words, one may expect that in this limit the term $A\, c\, S  \exp(-a T)$ also must disappear. This suggests that the backreaction coefficient $c$ should be also suppressed by the warping factor $e^{2 {\cal {A}}_{0} } \ll 1$, which makes the assumption $b\ll A\, c$ unwarranted.

This suggestion receives a confirmation from the consistency analysis of the model  \rf{KWmix}.  As we are going to show in the next section,  the model in \rf{KWmix} satisfies the consistency condition required in the theory with a nilpotent multiplet $S$ only if $|A\, c|\leq |b|$, which exactly matches our expectations. Note that the   term $A\, c\,  \exp(-a T)$ is exponentially suppressed as compared to the term $A\, c\,$, so for $|A\, c|\leq |b|$ we return back to the standard KKLT model  with consistent uplifting described by the theory \rf{KKLT}.

One may try to leave the consistency  issues aside and interpret the scenario described by \rf{KWmix} as a model of a heavy stabilized chiral field $S$ with a coupling $b \, S + A\, c\, S \exp(-a T)$. This will break the supergravity interpretation of the $\overline{\rm D3}$ brane uplifting in KKLT, but it will not affect the presence of metastable dS minima in the scalar potential, even in the case $b\ll A\ c$. The reason is that the term $A\, c\,  \exp(-a T)$ becomes exponentially suppressed at large $T$, whereas the coefficient  $b$ remains constant. As a result, the positions of the dS minima at large $c$ are proportional to $\log{1/b}$. Therefore, these minima effectively disappear once we take  $b=0$, as it was considered in \cite{Moritz:2017xto}.  This only indicates that this limit does not capture the physics of the theory with small but non-vanishing $b$.

In the following, we will analyze in detail the arguments outlined above and draw our conclusions.

%=================================================================%
\section{Validity of the Nilpotency of $S$}\label{nil}
\label{sec:nilpotent}
%=================================================================%

An important property of the theory \rf{KKLT} is that it represents the so-called consistent de Sitter supergravity \cite{Bergshoeff:2015tra,Hasegawa:2015bza,Kallosh:2015sea,Kallosh:2015tea,Schillo:2015ssx,Freedman:2017obq}
with the nilpotent multiplet interacting with standard chiral multiplets with non-linearly realized  supersymmetry. This is a local version of the Volkov-Akulov supersymmetry    which has only fermions in the multiplet\cite{Volkov:1972jx}.

A necessary condition for the consistency of de Sitter supergravity in \cite{Bergshoeff:2015tra,Hasegawa:2015bza,Kallosh:2015sea,Kallosh:2015tea,Schillo:2015ssx}   is that the term linear in $S$ in the superpotential is a non-vanishing function of other moduli, {\it e.g.} for $W(T,S)= W(T) + S f(T)$ one has
\be
D_S\, W(T,S) \big |_{S=0}= f(T) \neq 0 \ .
\ee
The reason for this is that there are vertices in the Lagrangian  with non-linearly realized supersymmetry, which have couplings ${1\over f(T)}$, see \cite{Bergshoeff:2015tra,Hasegawa:2015bza,Kallosh:2015sea,Kallosh:2015tea,Schillo:2015ssx,Freedman:2017obq}. The off-shell action, which has a non-linearly realized supersymmetry, must contain non-vanishing $f(T)$, for the consistent supersymmetric embedding to exist.

At the level of the components $(s, \psi, F)$ of the superfield $S$, the nilpotency condition $S^2=0$ translates into 3 equations
\be
s^2(x)=0, \qquad s(x) \psi=0,  \qquad 2 s (x) F - \psi^2=0 \ .
\ee
If $F=D_SW\neq 0$ we find that there is a Volkov-Akulov fermionic field, whereas the scalar degree of freedom becomes replaced by a bilinear fermion such as
\be
s=\psi\psi/(2 F) \ .
\ee
If $F=D_SW=0$ there is only the trivial  solution
\be
s(x)=\psi(x) = F=0 \ .
\ee
At the level of the connection with the $\overline{\rm D3}$ brane, this means that the effective tension vanishes and one cannot assume its existence at that point. 

In the KKLT model \rf{KKLT} one finds that, everywhere in the moduli space of $T$, the existence of the $\overline{\rm D3}$ is equivalent to the nilpotency condition on the $S$ superfield
\be
D_S\, W\big |_{S=0}^{c=0}= b\neq 0 \ .
\ee
The supergravity action with non-linearly realized supersymmetry is thus well defined for any value of $T$.

Let us now consider the model \eqref{KWmix} with the new coupling term proportional to $c$. According to \cite{Moritz:2017xto}, this would represent some unknown non-perturbative correction to the superpotential,  which depends on the anti-brane tension.  In this case, the supersymmetry breaking in the $S$ direction reads
\begin{equation}
D_S W |_{S=0}^{c\neq 0} = b+c\ A  e^{-a T}\,,
\end{equation}
thus showing an explicit dependence on the field $T$. This implies that there exists a point in the complex plane of moduli space where supersymmetry is restored in the $S$ direction, that is 
\begin{equation}\label{sing}
 D_S W = 0  \qquad \Rightarrow \qquad T_0= \frac{1}{a}\ln\left(-\frac{cA}{b}\right)\, .
\end{equation}
 For positive $A$, the position of $T_0$  depends on the relative signs of $b$ and $c$. Specifically, it is located on the real axis $\text{Im}\,T=0$ when $b\cdot c <0$, while this shifts to $\text{Im}\,T=\pm \pi/a$ for $b\cdot c >0$.

However a point in field space with $D_S W = 0$ is certainly problematic with the assumption of  \cite{Moritz:2017xto} that the field $S$ is nilpotent. Specifically, as explained above, the existence of this point is inconsistent with the  known version of the dS supergravity action \cite{Bergshoeff:2015tra,Hasegawa:2015bza,Kallosh:2015sea,Kallosh:2015tea,Schillo:2015ssx,Freedman:2017obq} and we do not know if any other version of it, allowing vanishing $D_SW$, will be ever constructed. Thus, the addition of a term $c\, A\ S e^{-a T} $ to the KKLT superpotential means that the corresponding model, at some point of the moduli space where $D_SW=0$, does not have an $\overline{\rm D3}$ brane. Technically, the potential in \rf{KWmix}  does not have a supersymmetric embedding.

 One could hope that there is some way to overcome this problem and generalize  dS supergravity action  \cite{Bergshoeff:2015tra,Hasegawa:2015bza,Kallosh:2015sea,Kallosh:2015tea,Schillo:2015ssx,Freedman:2017obq}  to make is compatible with the  singularity of the coupling, but it does not seem likely. In any case, until it is done, the modification of the KKLT
superpotential 
proposed in  \cite{Moritz:2017xto} has some clear limitations. By looking at eq.~\eqref{sing} one finds that the only  possibility  to consistently use the model \eqref{KWmix} is to impose the constraint
\be
|A\, c|\leq|b|\ .
\ee
This confines the `critical point' $T_0$ to $T+\bar T <0$, which is outside the half-plane $T+\bar T >0$ where supergravity with the \K\ potential \rf{KKLT}, \rf{KWmix} is defined. In that case, the superfield $S$ is nilpotent in the whole  allowed field space $T+\bar T >0$. However, for $|A\, c|\leq|b|$, the new contribution $c\, A\, S\, e^{-a T}$ is exponentially smaller than the original term $b\,  S$, and  physical consequences of that model, including moduli stabilization and uplifting,  become identical to the ones typical of KKLT and described by \eqref{KKLT}.

%=================================================================%
\section{On the  absence of \NoCaseChange{d}S vacua in KKLT according to \cite{Moritz:2017xto}}
\label{sec:dS}
%=================================================================%
\vspace{-3mm}

Ignoring  the supersymmetry and nilpotency issues, which by itself are sufficient to disqualify the model in \rf{KWmix}, we 
can still check whether the analysis of the bosonic potential 
 based on  \rf{KWmix} agrees with our findings. We will  reach a conclusion opposite to the one in \cite{Moritz:2017xto}.

The argument in  \cite{Moritz:2017xto} as to why there are no dS vacua in \rf{KWmix}  is explained at p. 30, after the analysis in d=10. We quote: ``It is now interesting to go back to the 4D superpotential (17) to note that our 10D analysis is compatible with the extreme case $b \rightarrow 0$, suggesting the existence and significant strength of the superpotential term describing the interaction between the anti-brane and the gaugino condensate (see Fig. 4)''. The reader is invited to  look  at   Fig. 4 where the potential is plotted at $b=0$ and it is explained why this potential cannot be uplifted.

Let us now stress that in the d=10 analysis in \cite{Moritz:2017xto} it was never proven that $b=0$, all d=10  arguments were given up to some unknown factors. At best the d=10 argument might suggest  that $b$ is small. Therefore we will now compare the properties of the scalar potential following from  \rf{KWmix} at small $b$ versus the one at $b=0$.

\vspace{-2mm}

%=================================================================%
\section{de Sitter vacua }\label{sds}
\label{sec:modstab}
%=================================================================%

\vspace{-3mm}

The scalar potential corresponding to the theory \eqref{KWmix}  is given by
\bea
V= {e^{-2 a t}\over 12 t^2} &&\Big(A^{2} c^{2} + 2A\, b\, c\, e^{a t} \cos  a \theta +b^{2}e^{2a t} + 2 a^2 A^2 t  \nonumber \\
+ &&6 a A (A +W_{0} e^{a t}  \cos  a \theta )\Big) \ \label{v},
\eea
where $T = t+i\theta$. This potential has extrema with respect to $\theta$ at $\theta = 0$, or at $\theta =  \pi/a$, modulo  $2\pi n/a$.  The stability condition along these two different directions imposes the combination $bc +3aW_0$ to be negative or positive, respectively. Note that, in the original case of KKLT, this option is absent, since for $c=0$  the only stable direction, in the fundamental domain, is the real axis.

The main argument given in  \cite{Moritz:2017xto} is that, if  $b$ is many orders of magnitude smaller than $c$, one can simply ignore $b$ and consider the potential \rf{v} in the limit $b = 0$. This yields the potential shown in Fig.~4 of  \cite{Moritz:2017xto}. From that figure, it is obvious that the potential at $b = 0$ does not describe dS vacua for any $c$.

While this argument may intuitively seem appealing, it is actually incorrect, because in \rf{v}  the small constant $b$ is amplified by the exponentially large factor $e^{a t}$. 

\subsection{\boldmath{dS vacua at $T = t + i {\pi\over a}$}}

We will start our analysis with the   potential \rf{v} for $\theta =  \pi/a$, since this analysis is quite simple. The potential for $T = t+ i\,\pi/a$  is given by
\be\label{v2}
V= {e^{-2 a t}\over 12 t^2}  \big((A c - b e^{a t})^2 + 2 a^2 A^2 t +  6 a A (A - W_0e^{a t})\big)  .
\ee
An identical potential describes the theory at $\theta = 0$ if one simultaneously changes the sign of $c$ and $W_0$ in \rf{v}.

First of all, let us notice that the potential \rf{v2} is manifestly positive, because the value of $W_0$ used in the original version of the KKLT scenario is negative. Therefore if this potential has any minimum, it is definitely dS.

Secondly, let us consider the large $c$ limit. Then one can easily understand that at small $t$, as well as at large $t$, the main contribution to the potential is given by the term
\be
{e^{-2 a t}\over 12 t^2}  \big(A c - b e^{a t}\big)^2 \ .
\ee
This term dominates everywhere outside a small vicinity of the point $b e^{a t}=A c$, where it vanishes. Therefore one could argue that at large $c$ this potential should have a dS minimum at 
\be\label{crp}
t \sim {1\over a} \log {Ac\over b}\ , \qquad \theta = {\pi\over a} \ ,
\ee
i.e. close to the critical point \rf{sing}.

\begin{figure}[!ht]
%\vspace*{3mm}
\hspace{-3mm}
\begin{center}
\includegraphics[width=7.5cm]{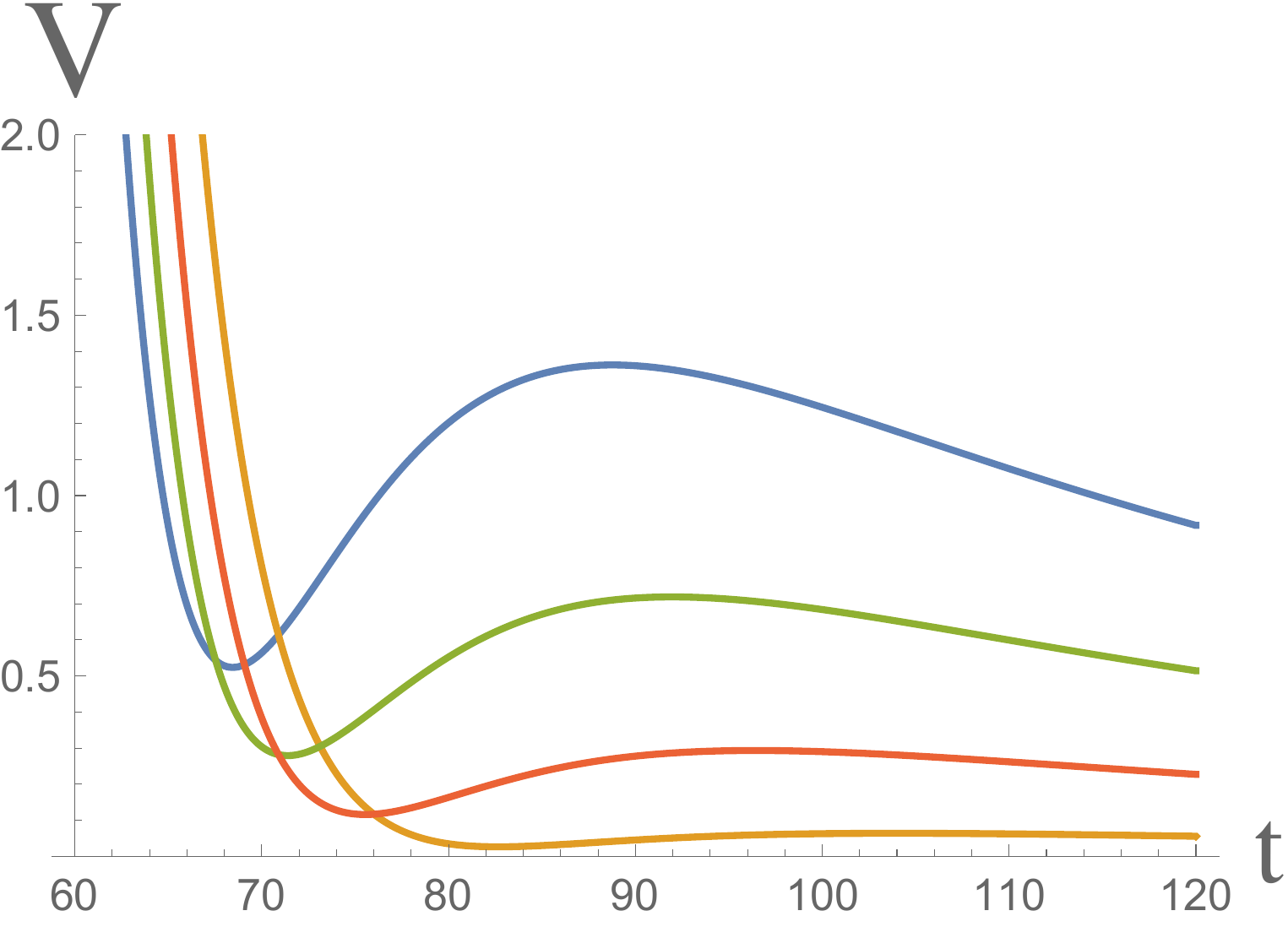}
\caption{The potential \rf{v2} (multiplied by $10^{{10}}$) for  $A=1$, $a=0.1$, $W_0=-10^{-4}$, as in the original version of the KKLT model. As an illustration, we take $c = 3$ and plot the potential for  $b=0.001$, $0.002$, $ 0.003$, and $0.004$, from left to right.}
\label{f1}
\end{center}
\vspace{0cm}
\end{figure}

To confirm this conclusion, we plot the potential \rf{v2} for  $A=1$, $a=0.1$, $W_0=-10^{-4}$, $c= 3$ and  \mbox{$b=\{0.001,\  0.002,\ 0.003,\ 0.004\}$}. As we see,  the potential has a dS minimum for each of these values of $b$.  We found that each of these  dS minima appear close to the position of the critical point \rf{crp} for the corresponding values of the parameters, and accuracy of this estimate become better and better at large $c$ and small $b$.

These considerations clearly show that, in the context of the KKLT stabilization, one should be very careful with considering apparently irrelevant terms. In the situations with $b \ll c$, the two contributions become nevertheless comparable and one cannot drop one of the two in the analysis of the model. Eq. \rf{crp} shows that the theory has dS vacua  for extremely small values of $b$, contrary to the expectations of   \cite{Moritz:2017xto}. Moreover, the position of the minimum, which appears to be close to  $t = {1\over a} \log {Ac\over b}$, runs to infinity in the limit $b \to 0$.

One should note that, since the potential \rf{v2} is manifestly positive, one must  take exponentially small $b$  in order to describe the cosmological constant $\Lambda  \sim 10^{{-120}}$. 

Now we will turn to the potential \rf{v1}, which is more closely related to the original KKLT potential and allows to get $\Lambda  \sim 10^{{-120}}$ in the context of the string landscape scenario without using extreme values of parameters.

\subsection{\boldmath{dS vacua at  $T = t $}}

The standard KKLT potential has a minimum at  real $T$ with $\theta = 0$. Looking at the same direction,  the modified potential used in \cite{Moritz:2017xto} is given by
\be\label{v1}
V= {e^{-2 a t}\over 12 t^2}  \big((A c + b e^{a t})^2 + 2 a^2 A^2 t +  6 a A (A + W_0 e^{a t} )\big)  .
\ee
\begin{figure}[!ht]
%\vspace*{3mm}
%\hspace{-3mm}
\begin{center}
\includegraphics[width=8.3cm]{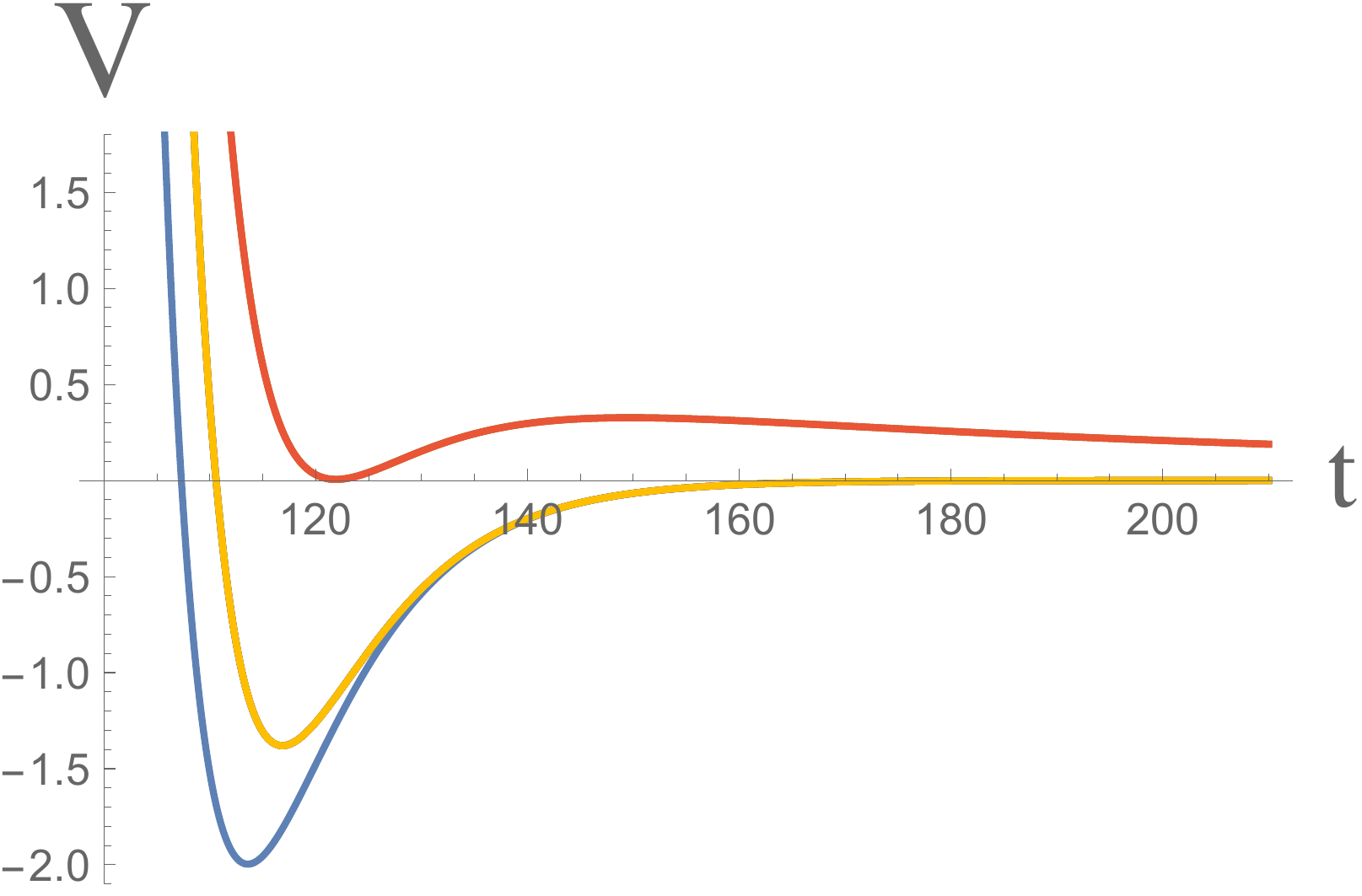}
\caption{The potential \rf{v1} (multiplied by $10^{15}$) for  $A=1$, $a=0.1$, $W_0=-10^{-4}$. The blue (lower) line shows the potential with a supersymmetric AdS minimum prior to uplifting, at $b = c = 0$. The second (yellow),  line shows the potential at $b = 0$ uplifted by increase of $c$ to  $c = 1$. This does not uplift the potential to dS. Finally, the upper (red) line shows the potential with a dS (nearly Minkowski) minimum for $c = 1$, $b =10^{{-5}}$. The main part of the uplifting is not due to the large change of $c$ from 0 to 1, but due to the tiny change of $b$ from 0 to $b =10^{{-5}}$.}
\label{f2}
\end{center}
\vspace{0cm}
\end{figure}

Analytical investigation of the potential \rf{v1} is somewhat more involved than the investigation of the potential \rf{v2}, but the final conclusion of our numerical analysis is very similar. At large $c$, this potential has dS minima at a position almost exactly coinciding with $t \sim {1\over a} \log {Ac\over b}$, see Fig. \ref{f2}. 
We plotted the potential for $c = 1$, but dS potential exist for much greater and for much smaller values of the parameter $c$. Such potentials can describe metastable  dS vacua with arbitrarily small values of the cosmological constant, including $\Lambda \sim 10^{{-120}}$.
\begin{figure}[!ht]
%\vspace*{3mm}
\begin{center}
{\hspace{-3mm}\includegraphics[width=8.8cm]{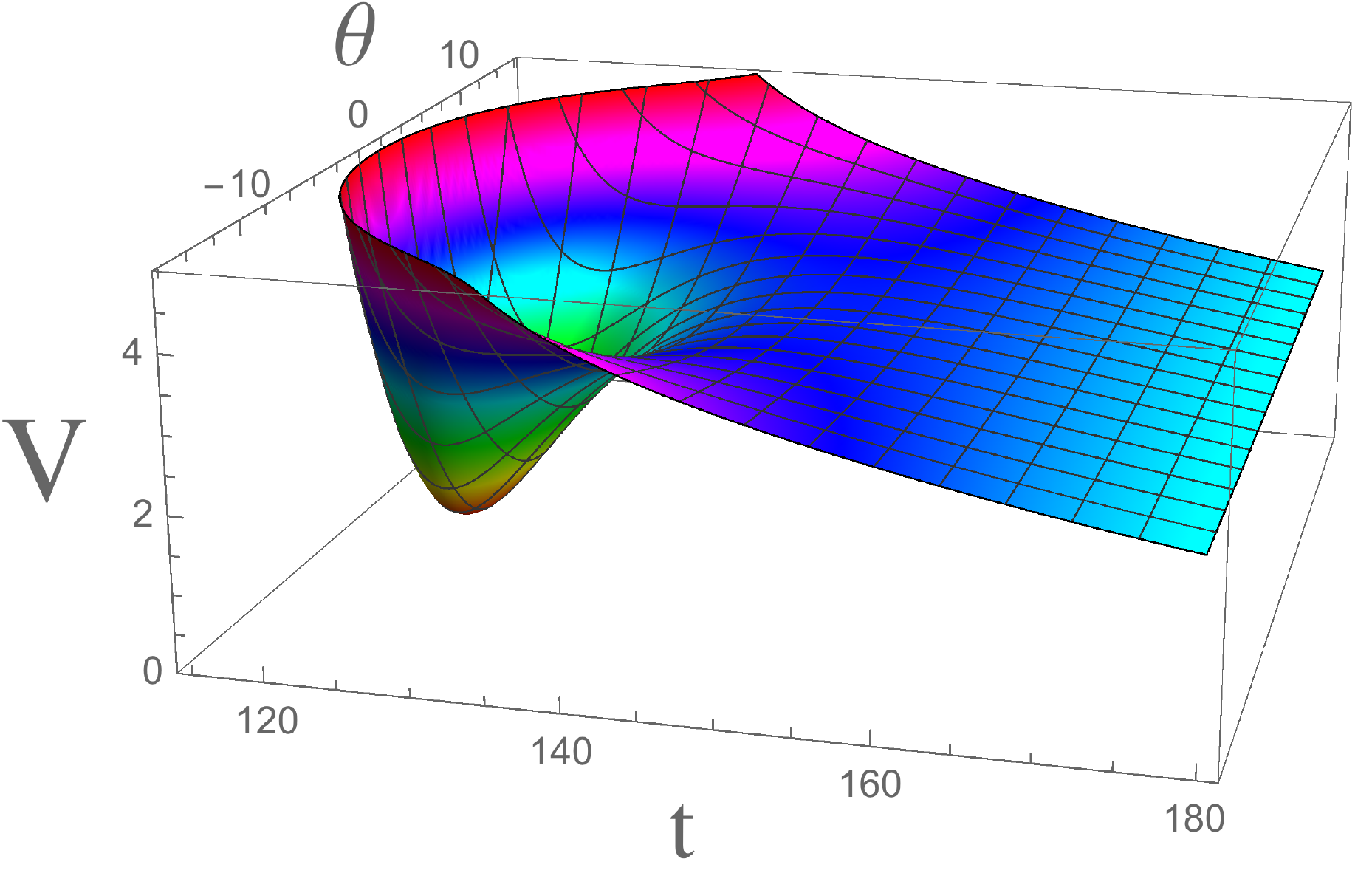}}
\caption{The potential \rf{v1} (multiplied by $10^{{16}}$) for  $A=1$, $a=0.1$, $W_0=-10^{-4}$, $c = 1$ and $b = 10^{{-5}}$. The potential has a dS minimum with a tiny cosmological constant.}
\label{f3}
\end{center}
\vspace{0cm}
\end{figure}
One can check that these dS minima (just as the minima studied in the previous subsection) are indeed true minima with respect to both of the fields $t$ and $\theta$. We illustrate it in Fig. \ref{f3}, which shows the potential \rf{v1} as a function of $t$ and $\theta$   for  $A=1$, $a=0.1$, $W_0=-10^{-4}$, $c = 1$ for $b = 10^{{-5}}$. The potential has a dS minimum with a tiny positive cosmological constant.

Returning to Fig. \ref{f2}, this figure reveals some instructive facts. First of all, contrary to the arguments of  \cite{Moritz:2017xto}, during the full process of uplifting the position of the minimum is shifted only by 7.3\%. Also, the shift from $c = b = 0$ to $c =1$, $b = 0$, uplifts the minimum from -2 to -1.6, i.e. only by 1/5 of the way from the supersymmetric AdS to dS. The main part of  uplifting is achieved due to the tiny change from $b = 0$ to $b =10^{-5}$.  This shows once again how dangerous it is to ignore string tension encoded in $b$ as compared with the backreaction $c$, even if one assumes that it makes sense to consider large backreaction on the tension of the $\overline{\rm D3}$ brane in the limit when this tension vanishes, see a discussion of this issue in the Introduction.

\begin{figure}[!h]
%\vspace*{3mm}
\begin{center}
{\hspace{-3mm}\includegraphics[width=7.5cm]{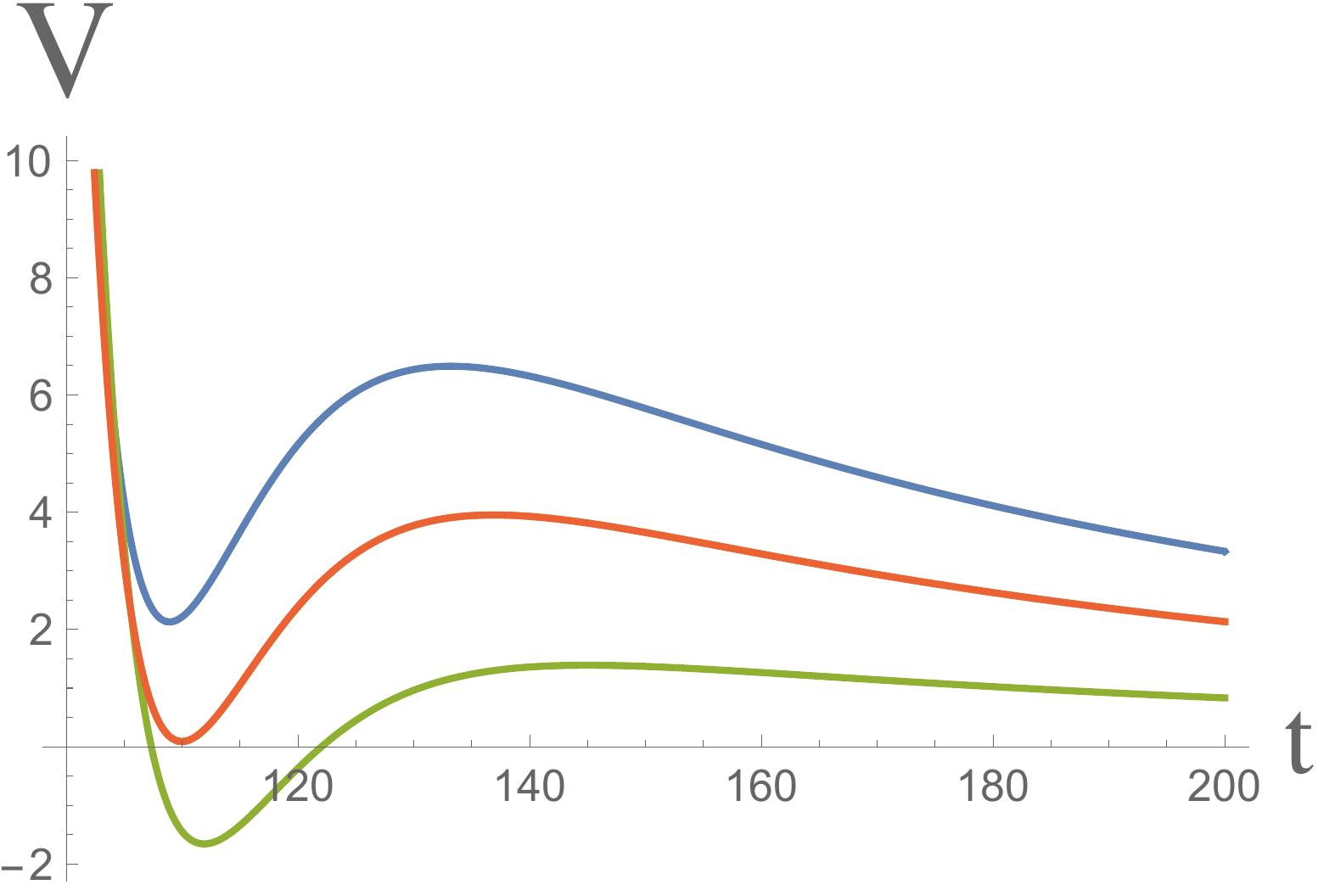}}
\caption{The potential \rf{v1} (multiplied by $10^{{15}}$) for  $A=1$, $a=0.1$, $W_0=-10^{-4}$, $c = -1$ and $b = 2\times 10^{{-4}}$, $3.2\times 10^{{-4}}$, and $4\times 10^{{-4}}$. The potential with $3.2\times 10^{{-4}}$ has a dS minimum with a tiny cosmological constant.}
\label{f4}
\end{center}
%\vspace{0.3cm}
\end{figure}

For completeness, one may also consider models with $c < 0$. In Fig. \ref{f4} we show a family of dS vacua  for $c = -1$, $b = 2\times 10^{{-4}}$, $3.2\times 10^{{-4}}$, and $4\times 10^{{-4}}$. 
\vspace{0.5cm}

%=================================================================%
\section{From 10\NoCaseChange{d} to 4\NoCaseChange{d}}
\label{sec:10d}
%=================================================================%

As we can see,  the results of our investigation of the modified 4d KKLT model \rf{KWmix} proposed in  \cite{Moritz:2017xto} disagree  with the statement that KKLT construction fails to uplift based on the  10d analysis performed in \cite{Moritz:2017xto}.   
 
 The central point of the 10d argument is an {\it assumption} near eq.~(62) which says ``We simply assume that the new minimum lies at a different volume'', $\rho\rightarrow \rho +\delta \rho$. Concerning the actual value of $\delta \rho$ proposed in eq. (63), they explain that they cannot compute its value.  
 
The main computation related to the  10d analysis in   \cite{Moritz:2017xto} is in Appendix C. It basically re-derives the case of the backreaction due to mobile D3 branes  placed on  D7 branes that wrap 4-cycles of the Calabi-Yau, as it was done before in \cite{Dymarsky:2010mf}. 
 There is no such computation for the $\overline{\rm D3}$ at the tip of the conifold neither in \cite{Dymarsky:2010mf}, nor in  \cite{Moritz:2017xto}. Further, any analogy of the mobile D3 brane analysis in \cite{Dymarsky:2010mf} with \cite{Kallosh:2014wsa,Bergshoeff:2015jxa} is complicated by the fact that \cite{Kallosh:2014wsa,Bergshoeff:2015jxa} explicitly considered an $\overline{\rm D3}$ placed on an orientifold plane; the latter of which is fixed in position\footnote{This is strictly true only in perturbation theory in the string coupling $g_s$, and at weak coupling, which we  restrict ourselves to here.  Making progress beyond this requires the use of F- and M-theory.} and hence is in no sense mobile. Instead, in    \cite{Moritz:2017xto}  they give a ``parametric estimate (neglecting volume powers) based on generic assumptions'' and end up suggesting that $\delta \rho$ is significant.\footnote{An intuitive argument provided to us by E. Silverstein and \mbox{S. Kachru} is that the backreaction of the anti-D3 brane sitting far away at the end of the throat can only be small, which is why they believe that $c$ in the 4d superpotential \rf{KWmix} makes sense only if $c$ is small, in agreement with our nilpotent field analysis. In this case the standard KKLT uplift does take place. } These assumptions were the basis for neglecting the coefficient $b$ representing the tension of the anti-D3 brane.  The final results of the 10d analysis are illustrated in  \cite{Moritz:2017xto} by Fig. 4 demonstrating the absence of uplifting in the 4d model for $b = 0$.  Meanwhile our calculations show that even if $b$ is 5 orders of magnitude smaller than  $c$, the shift  $\delta \rho$ during uplifting is very small,  and dS vacua do exist,  see Fig.~\ref{f2}.
  
In short, we note that the assumptions made in the context of the 10d analysis in \cite{Moritz:2017xto} follow a very similar trend of the assumptions made in the 4d model (or vice-versa), which we already proved to be  incorrect.  The intuition behind the 10d investigation about the significant backreaction of the anti-D3 brane sitting at the end of the throat  {\it by analogy}  with the backreaction due to D3-brane positions on D7-branes wrapped on 4-cycles, is actually not supported by explicit computations and contradicts the general arguments given in the Introduction of this paper. 

This leads us to suggest  that in fact there should be an equivalence between our 4d and a (properly corrected) 10d analysis. This investigation is the subject of \cite{Kallosh:2018nrk}, where a  derivation of de Sitter supergravity in d=4 from a string compactification in d=10 is proposed. Further discussion of the 10d physics can be found in \cite{Akrami:2018ylq}, \cite{Cicoli:2018kdo}, and \cite{Kachru:2018aqn}.

%=================================================================%
\section{Conclusion}
\label{sec:conclusion}
%=================================================================%

As we have shown in Section \ref{nil}, the modification of the KKLT model proposed  in  \cite{Moritz:2017xto} and shown here in our equation \rf{KWmix}, is problematic because it violates the nilpotency condition at some point in the moduli space. The corresponding potential does not have a supersymmetric embedding in the class of  currently available supergravities with nilpotent multiplet  \cite{Bergshoeff:2015tra,Hasegawa:2015bza,Kallosh:2015sea,Kallosh:2015tea,Schillo:2015ssx,Freedman:2017obq}, unless $|A\, c|\leq|b|$, in which case the standard KKLT uplifting scenario takes place.

Secondly, and independently from the SUSY embedding and nilpotency issues, the   analysis of the bosonic potential in  \cite{Moritz:2017xto} was based on an assumption that the small parameter $b$ in the superpotential can be safely taken as  $b=0$, despite the fact that the other term in the superpotential, proportional to $c$,  decreases exponentially at large $T$. In Section \ref{sds}   we have shown   that  dS minima do actually exist for a broad range of parameters $c$ and $b$, with an exception of the unphysical limiting case $b = 0$ considered in  \cite{Moritz:2017xto}. 
Therefore we conclude  that the critical arguments against the KKLT construction in  \cite{Moritz:2017xto} based on the 4d supergravity approach are not valid, whereas the 10d arguments are debatable.
 
  \vspace{-.5cm}

\acknowledgments
 \vspace{-.5cm}
We would like to thank Stephon Alexander, Jose J. Blanco-Pillado, Fotis Farakos, Arthur Hebecker, Shamit Kachru, Fernando Quevedo,  Ander Retolaza, Eva Silverstein, Sandip Trivedi, Thomas Van Riet, Alexander Westphal, and Timm Wrase for helpful comments and discussions. The work  of RK and AL is supported by SITP,  by the NSF Grant PHY-1720397, and by the Simons Foundation grant.  EM is supported in part by the National Science and Engineering Research Council of Canada via a PDF fellowship.  MS is supported by the Research Foundation - Flanders (FWO) and the European Union's Horizon 2020 research and innovation programme under the Marie Sk{\l}odowska-Curie grant agreement No. 665501.  MS acknowledges also financial support by the foundation `Angelo Della Riccia' for his research stay at DESY, where this work was initiated. 
%

%%%%%%%%%%%%%%%%%%%%%%%%%%%%%%%%
%%%%%%%%%%%%Bibliography%%%%%%%%%%%%%%
%%%%%%%%%%%%%%%%%%%%%%%%%%%%%%%%
\bibliography{lindekalloshrefs}
%\bibliography{refsN}
\bibliographystyle{utphys}

\end{document}